\documentclass[mathleft
]{an}
\usepackage{graphicx}
\usepackage{times}
\usepackage{amsmath}
\newcommand{\ovec}[1]{{\mbox{\boldmath $#1$}}}

\newcommand{\ba}{\ovec{a}}
\newcommand{\bA}{\ovec{A}}
\newcommand{\bb}{\ovec{b}}
\newcommand{\bB}{\ovec{B}}

\newcommand{\be}{\ovec{e}}
\newcommand{\bE}{\ovec{E}}
\newcommand{\bscE}{\ovec{\cal{E}}}

\newcommand{\bF}{\ovec{F}}
\newcommand{\bscF}{\ovec{\cal{F}}}
\newcommand{\bg}{\ovec{g}}

\newcommand{\bj}{\ovec{j}}
\newcommand{\bJ}{\ovec{J}}
\newcommand{\bk}{\ovec{k}}

\newcommand{\bu}{\ovec{u}}
\newcommand{\bU}{\ovec{U}}

\newcommand{\bx}{\ovec{x}}

\newcommand{\bepsilon}{\ovec{\epsilon}}

\newcommand{\bxi}{\ovec{\xi}}

\newcommand{\bOmega}{\ovec{\Omega}}

\newcommand{\bmB}{\overline{\ovec{B}}}
\newcommand{\bmE}{\overline{\ovec{E}}}
\newcommand{\bmF}{\overline{\ovec{F}}}
\newcommand{\bmJ}{\overline{\ovec{J}}}
\newcommand{\bmU}{\overline{\ovec{U}}}
\newcommand{\nab}{\ovec{\nabla}}
\newcommand{\cE}{\cal{E}}

\newcommand{\mB}{\overline{B}}
\newcommand{\mF}{\overline{F}}
\newcommand{\mG}{\overline{G}}
\newcommand{\x}{\times}
\newcommand{\p}{\partial}
\newcommand{\dd}{\mbox{d}}
\newcommand{\iu}{\mbox{i}}
\newcommand{\ol}{\overline}
\newcommand{\bzo}{{\bf 0}}
\newcommand{\EQ}{\begin{equation}}
\newcommand{\EN}{\end{equation}}
\newcommand{\EQA}{\begin{eqnarray}}
\newcommand{\ENA}{\end{eqnarray}}
\newcommand{\eq}[1]{(\ref{#1})}

\begin{document}

% The following seven commands are intended for editorial usage and should be ignored by
% the author(s).
\Pagespan{789}{}% Document's page range.
% If second parameter is left empty, the last page is computed automatically.
\Yearpublication{2014}%
\Yearsubmission{2014}%
\Month{01}%
\Volume{XXX}%
\Issue{YY}%
% \DOI{This.is/not.aDOI}%

\title{Mean-field dynamos:\\
       the old concept and some recent developments}

\author{K.-H. R\"adler\thanks{\email{khraedler@arcor.de}\newline}}

\titlerunning{Mean-field dynamos}
\authorrunning{K.-H. R\"adler}

\institute{
Leibniz-Institut f\"ur Astrophysik, An der Sternwarte 16,
D-14482 Potsdam, Germany
}

\received{XX January 2014}
\accepted{YY ZZZ 2014}
\publonline{later}

\keywords{Cosmic magnetic fields -- mean-field electrodynamics -- mean-field magnetohydrodynamics -- mean-field dynamos}

\abstract{%
This article reproduces the Karl Schwarzschild lecture 2013.
Some of the basic ideas of electrodynamics and magnetohydrodynamics of mean fields in turbulently moving conducting fluids are explained.
It is stressed that the connection of the mean electromotive force with the mean magnetic field and its first spatial derivatives
is in general neither local nor instantaneous and that quite a few claims concerning pretended failures of the mean-field concept
result from ignoring this aspect.
In addition to the mean-field dynamo mechanisms of $\alpha^2$ and $\alpha \Omega$ type several others are considered.
Much progress in mean-field electrodynamics and magnetohydrodynamics results from the test-field method for calculating
the coefficients that determine the connection of the mean electromotive force with the mean magnetic field.
As an important example the memory effect in homogeneous isotropic turbulence is explained.
In magnetohydrodynamic turbulence there is the possibility of a mean electromotive force that is primarily independent
of the mean magnetic field and labeled as Yoshizawa effect.
Despite of many efforts there is so far no convincing comprehensive theory of $\alpha$ quenching, that is, the reduction
of the $\alpha$ effect with growing mean magnetic field, and of the saturation of mean-field dynamos.
Steps toward such a theory are explained.
Finally, some remarks on laboratory experiments with dynamos are made.
}

\maketitle

\section{Introduction}

At the beginning of the last century mankind knew the magnetic field of the Earth,
but nothing about magnetic fields at other celestial bodies.
In 1908 George Ellery Hale proposed to interpret line splittings in the spectrum of the light
coming from sunspots, which were not understood at this time, as a consequence of strong magnetic fields
(of a few kilogauss) within them.
Eleven years later, in 1919, Sir Joseph Larmor came up with the idea that magnetic fields at the Sun
could be generated by self-exciting dynamos just as introduced in engineering for instance
by Ernst Werner von Siemens in 1867
\footnote{The idea of the self-exciting dynamo has been stated several years before by Anyos Jedlik,
by S{\o}ren Hjorth and by Samuel Alfred Varley, in 1867 also by Charles Wheatstone.
Von Siemens is known for having recognized the practical importance of the dynamo principle.}.
Of course, Larmor's proposal was not readily accepted and there were many attempts to check it or to rule it out.
A dynamo in a homogeneous fluid is quite different from its technical version built up with insolated wires.
More mathematically spoken, a dynamo working in a singly-connected conducting region is different
from that in a multiply-connected region.
In 1934 Thomas George Cowling proved a theorem which we may now (after some generalizations) formulate by saying
that a dynamo can never work with an axisymmetric magnetic field.
Another important theorem traces back to investigations by Walter M. Elsasser in 1946
and by Edward C. Bullard and H. Gellman in 1954
and excludes a dynamo in a sphere due to motions without radial components.
Quite a few modifications of such theorems have been proven in the course of time showing the impossibility of dynamos
with some simple geometrical structures of the magnetic field or the fluid flow.

In 1947 Horace W. Babcock discovered a star with a strong magnetic field (of 34 kilogauss),
and in 1958 he published a catalogue of magnetic stars.
Later we learned about a large number of various objects which exhibit magnetic fields, including galaxies
with rather weak but very extended fields (of the order of $10^{-5}$ gauss) or neutron stars with extremely strong ones
(up to the order of $10^{15}$ gauss).

A rigorous existence proof for a homogeneous dynamo has been delivered by A. Herzenberg in 1958.
The velocity distribution he assumed was, however, far away from from that expected in the Earth's interior,
in the Sun or in stars.

Already before Herzenberg's proof, 1955 and 1957, Eugene N. Parker designed a model for the Sun in which
``cyclonic convection" together with rotational shear, that is differential rotation, produce a magnetic cycle.
In 1964 Stanislaus I. Braginsky published his model of the ``nearly symmetric dynamo" which reflects features
of the Earth's magnetic field.

In the early sixties Max Steenbeck in Jena pushed Fritz Krause and myself to think about the question of how
the Sun or the Earth could generate their magnetic fields.
Many conceivable mechanisms were discussed and investigated in the course of time.
At the end the mean-field electrodynamics of electrically conducting turbulently moving fluids was established.
A central issue of this theory is the $\alpha$ effect, the occurrence of an electromotive force with a part parallel
(or antiparallel) to the mean magnetic field as a consequence of induction processes caused by irregular motions.
The $\alpha$ effect allows dynamo action.
The first paper on this topic has been published by Steenbeck, Krause and R\"adler in 1966
(unfortunately only in German language).
Since then mean-field electrodynamics and, more general, mean-field magnetohydrodynamics have been elaborated in great detail
and dynamo models have been proposed for the Sun, planets, several types of stars and for galaxies.
In this lecture I would like to explain a few results of this research field.
(For comprehensive presentations see, e.g., Moffatt 1978, Krause and R\"adler 1980 or Brandenburg and Subramanian 2005.)

Before doing so, however, I would like to say:
It is a great honor for me to receive the Karl Schwarzschild Medal.
I am very grateful to the Board of the Astronomische Gesellschaft for this distinction.
It is also a great honor to deliver this Schwarzschild lecture.

Let me start with a few remarks about Max Steenbeck (1904-81) and the place, Jena,
where mean-field electrodynamics was born.
Max Steenbeck was no geo- or astrophysicist.
He was one of the great pioneers of plasma physics, worked until the end of the Second World War
in the Siemens Company in Berlin, dealt there with heavy current technology, for example rectifiers,
constructed the first working betatron etc.
At the end of the war he has been interned in the Soviet Union.
He spent there (involuntarily) eleven years, dealing in particular with the separation of Uranium isotopes
in the framework of the Soviet Atomic program.
After his return to the G.D.R. he dealt there with magnetic materials, with nuclear power stations,
and in 1959  he established the Institute for Magnetohydrodynamics in Jena with the idea to deliver contributions
to nuclear fusion research, which looked at that time very promising.
Sometimes, on his frequent rides between Jena and occasionally on Saturdays and Sundays,
he thought about possibilities of processes in the Sun or in the Earth's interior that might produce
the observed magnetic fields, and then attacked Fritz Krause and me with his ideas.
Since the Institute for Magnetohydrodynamics was not a place with astrophysical or geophysical tradition
there was, at least at the beginning, no contact to leading scientists in these fields.

\section{Mean-field electrodynamics}

\subsection{The basic idea}

In what follows we deal with electromagnetic processes in an electrically conducting moving fluid.
Adopting the magnetohydrodynamic approximation we assume that the electromagnetic fields obey the pre-Maxwell equations
\EQ
\nab \x \bE = - \p_t \bB \, , \quad \nab \x \bB = \mu \bJ \, , \quad \nab \cdot \bB = 0
\label{eq01}
\EN
and Ohm's law for moving matter in the form
\EQ
\bJ = \sigma (\bE + \bU \x \bB) \, .
\label{eq03}
\EN
As usual, $\bE$ denotes the electric field, $\bB$ the magnetic field, $\bJ$ the electric current density,
and $\bU$ the fluid velocity; further $\mu$ means the magnetic permeability of free space
and $\sigma$ the electric conductivity of the fluid.
From \eq{eq01} and \eq{eq03} we may derive the induction equation
\EQ
\eta \nab^2 \bB + \nab \x (\bU \x \bB) - \p_t \bB = \bzo \, , \quad \nab \cdot \bB = 0 \, ,
\label{eq05}
\EN
with $\eta = 1 / \mu \sigma$ being the magnetic diffusivity.
For simplicity we ignore here any electromotive force independent of electromagnetic fields,
which would act as a battery.

Until further notice we consider the fluid velocity $\bU$ as given.
If the induction equation is solved and so the magnetic field $\bB$ is known,
we may calculate the electric field $\bE$ and the electric current density $\bJ$ without further integrations.

Thinking of the situation in many astrophysical objects, we assume further that the fluid velocity $\bU$ and so also
the electromagnetic fields $\bB$, $\bE$ and $\bJ$ show irregular fluctuations in space and time.
Considering these fluctuations we simply speak of turbulence (without having a specific definition of turbulence in mind).
We then focus attention on mean fields defined as averages of these fields and showing simpler dependencies
on space and time coordinates.
Space or time or statistical averages or combinations of them are admitted.
It is merely important that the Reynolds averaging rules, already known from hydrodynamic turbulence theory,
are (exactly or approximately) satisfied.
We denote averages by overbars.
Let $F$ and $G$ be quantities showing irregular behavior, that is fluctuations, and put $F = \mF +f$ and $G = \mG + g$.
Then these rules read
\EQA
&& \ol{F + G} = \mF + \mG
\label{eq05a}\\
&& \ol{\mF} = \mF \; \mbox{or, what is equivalent,} \; \ol{f} = 0
\label{eq05b}\\
&& \ol{F\,G} = \mF\,\mG + \ol{f\,g}
\label{eq05c}\\
&& \ol{\p F / \p x} = \p \mF / \p x \, , \quad \ol{\p F / \p t} = \p \mF / \p t \, .
\label{eq05d}
\ENA
We stress that the average of the product of two fluctuating quantities is not equal to the product
of the corresponding mean quantities, but there is an additional term determined by the fluctuations.

Returning to electrodynamics we subject equations \eq{eq01} and \eq{eq03} to averaging.
We find then their mean-field versions
\EQ
\nab \x \bmE = - \p_t \bmB \, , \quad \nab \x \bmB = \mu \bmJ \, , \quad \nab \cdot \bmB = 0
\label{eq11}
\EN
and
\EQ
\bmJ = \sigma (\bmE + \bmU \x \bmB + \bscE) \, .
\label{eq13}
\EN
When averaging \eq{eq05} in the same way we obtain
\EQ
\begin{aligned}
\!\!\!\! \eta \nab^2 \bmB + \nab \x (\bmU \x \bmB + \bscE) - \p_t \bmB &= \bzo \, ,\\
\qquad \qquad \qquad \qquad \qquad \nab \cdot \bmB &= 0 \, .
\end{aligned}
\label{eq15}
\EN
$\bscE$ is the mean electromotive force due to the fluctuations of the fluid velocity and the magnetic field,
$\bu = \bU - \bmU$ and $\bb = \bB - \bmB$, that is,
\EQ
\bscE = \ol{\bu \x \bb} \, .
\label{eq17}
\EN
The form of the equations \eq{eq11} - \eq{eq15} agrees widely with that of the original, not averaged equations \eq{eq01} - \eq{eq05}.
The only, but decisive deviation consists in the occurrence of the new electromotive force $\bscE$.

The crucial point in the elaboration of mean-field electrodynamics is the determination of that mean electromotive force $\bscE$.
We first consider $\bu$ as given.
As for $\bb$ we may derive from \eq{eq05} and \eq{eq15} that
\EQA
&& \!\!\!\!\!\!\!\!\!\!\!\! \eta \nab^2 \bb + \nab \x (\bmU \x \bb + \bepsilon) - \p_t \bb = - \nab \x (\bu \x \bmB)
\nonumber\\
&& \qquad \qquad \bepsilon = \bu \x \bb - \ol{\bu \x \bb} \, , \quad \nab \cdot \bb = 0 \, .
\label{eq19}
\ENA
Clearly $\bepsilon$ is the fluctuating part of $\bu \x \bb$.
Equation \eq{eq19} tells us that $\bb$ is a functional of $\bu$, $\bmU$ and $\bmB$, which is linear
(not necessarily linear and homogeneous) in $\bmB$.
Consequently, $\bscE$ depends also on these quantities and may be represented as a sum
\EQ
\bscE = \bscE^{(0)} + \bscE^{(B)}
\label{eq21}
\EN
of a part $\bscE^{(0)}$ independent of $\bmB$ and another part $\bscE^{(B)}$ which is is linear and homogeneous in $\bmB$.

Let us assume here that $\bb$ decays to zero if $\bmB$ vanishes.
This implies also the absence of small-scale dynamos (see section \ref{dynmech}).
Under this assumption $\bscE^{(0)}$ decays to zero, too.
As a result, $\bscE$ agrees with $\bscE^{(B)}$ and it must allow a representation in the form of the convolution
\EQ
\!{\cE}_i (\bx, t) = \!\!
\int_0^\infty \!\!\!\!\int_\infty \!\!\!\!  {\cal K}_{ij} (\bx,t; \bxi,\tau)  \mB_j (\bx+\bxi,t-\tau) \, \dd^3 \xi \, \dd \tau
\hspace{-1cm}
\label{eq25}
\EN
with some tensorial kernel ${\cal K}_{ij}$, which depends on $\bu$ and $\bmU$.
We know the explicit dependence of ${\cal K}_{ij}$ on $\bu$ and $\bmU$ only for very special cases,
but conclude from the turbulent nature of the velocity fluctuations that ${\cal K}_{ij}$ vanishes
for sufficiently large $|\bxi|$ and $\tau$.
As a consequence, $\bscE$ in a given point in space and time depends only on the behavior of $\bmB$
in a certain surroundings of this point, the extent of which is determined by the correlation length
and the correlation time of $\bu$.

It is appropriate to split the kernel ${\cal K}_{ij}$ in \eq{eq25} into two parts, one symmetric and the other one antisymmetric in $\bxi$,
and to express the last one by a derivative of a tensor symmetric in $\bxi$.
Doing so and subjecting then \eq{eq25} to an integration by parts we arrive easily at the equivalent representation
\EQ
\begin{aligned}
\!\!\!\!{\cE}_i (\bx,t) &= \int_0^\infty \!\!\! \int_\infty \!\! \Big( {\cal A}_{ij} (\bx,t; \bxi, \tau)
\mB_j (\bx + \bxi, t - \tau) \\
& \; + {\cal B}_{ijk} (\bx,t; \bxi, \tau) \frac{\p \mB_j (\bx + \bxi,t-\tau)}{\p x_k} \Big) \dd^3 \xi \dd \tau
\label{eq31}
\end{aligned}
\EN
with two tensors ${\cal A}_{ij}$ and ${\cal B}_{ijk}$ which are both symmetric in $\bxi$.

Assume for a moment that $\bmB$ varies only weakly in space and time, that is, there are distinct gaps in the spectra
of the length and time scales of the total magnetic field, $\bmB + \bb$,
separating large and small scales.
We speak then of ideal scale separation.
In this case relation \eq{eq31} turns into a simpler one,
\EQ
{\cE}_i = a_{ij} \mB_j + b_{ijk} \frac{\p \mB_j}{\p x_k}
\label{eq35}
\EN
with
\EQ
a_{ij} = \int_0^\infty \!\! \int_\infty \!\! {\cal A}_{ij} (\bx,t; \bxi,\tau) \, \dd^3 \xi \, \dd \tau
\label{eq37}
\EN
and an analogous connection between $b_{ijk}$ and ${\cal B}_{ijk}$.

While relation \eq{eq31} connects $\bscE$ at a given point in space and time with $\bmB$
in an arbitrary spatial surroundings of this point and at this time and arbitrary past times,
relation \eq{eq35} describes a local and instantaneous connection of $\bscE$ with $\bmB$
and its first spatial derivatives.
The latter relation, which has to be understood as application of the former one to idealized cases,
explains a large number of phenomena in turbulent fluids, in particular some types of dynamos.
It is, however, unable to capture, e.g., memory effects, that is, the dependence of the evolution of a mean field
not only on its current values but also on its history (see section \ref{imperfect}).
In what follows we will, as long as it is appropriate, refer to \eq{eq35} but switch to \eq{eq31} where necessary.
We want to stress that many statements on pretended failures of mean-field electrodynamics
or on allegedly narrow limits of its applicability result from ignoring \eq{eq31} and considering instead \eq{eq35}
as the most general relation for the mean electromotive force $\bscE$.

Let us finally mention the technical issue that convolutions like \eq{eq31} turn under a proper Fourier transformation
(or, concerning the time, also a Laplace transformation) into simpler algebraic relations.
Ignore for simplicity any dependence of ${\cal A}_{ij}$ and ${\cal B}_{ijl}$ on $\bx$ and $t$.
With a transformation
\EQ
\begin{aligned}
F (\bx,t) &= (2 \pi)^{-4} \int \!\! \int \hat{F} (\bk, \omega) \\
&\qquad \qquad \qquad \exp \big( \iu (\bk \cdot \bx - \omega t) \big)  \dd^3 k \, \dd \omega
\end{aligned}
\label{eq41}
\EN
relation \eq{eq31} turns then into
\EQ
\begin{aligned}
\hat{\cE}_i (\bk, \omega) &= \hat{\cal A}_{ij} (\bk, \omega) \, \hat{\mB}_j (\bk, \omega) \\
&\qquad \qquad \qquad  + \iu \, \hat{\cal B}_{ijk} (\bk, \omega) \, \hat{\mB}_j (\bk, \omega) k_k \, .
\end{aligned}
\label{eq43}
\EN

\subsection{A simple example}
\label{simple}

We consider first the case in which no mean flow exists, $\bmU = \bzo$,
and the velocity fluctuations $\bu$ correspond to a homogeneous isotropic turbulence.
As for the mean magnetic field $\bmB$ we assume ideal scale separation in the sense explained above
so that \eq{eq35} applies.

We define homogeneity of the turbulence by the invariance of all averaged quantities depending on $\bu$ under arbitrary translations
of the $\bu$ field, and isotropy by the invariance of all such quantities under arbitrary rotations of this field
about arbitrary axes.
We may also fix the $\bu$ field and subject the coordinate system in which we describe it to translations or rotations.
Then homogeneity and isotropy occur as invariance of all averaged quantities under arbitrary translations
and arbitrary rotations of the coordinate system.
In particular the tensors $a_{ij}$ and $b_{ijk}$ that occur in \eq{eq35} have to show these properties,
that is, their components have to be independent of space coordinates
and independent of rotations of the coordinate system.
This qualifies them as space-independent isotropic tensors, that is, they differ only by simple factors,
say $\alpha$ and $\beta$, from the Kronecker tensor $\delta_{ij}$ and the Levi-Civita tensor $\epsilon_{ijk}$,
that is, $a_{ij} = \alpha \, \delta_{ij}$ and $b_{ijk} = \beta \epsilon_{ijk}$.
In this way we arrive at
\EQ
\bscE = \alpha \, \bmB - \beta \, \nab \x \bmB \, ,
\label{eq51}
\EN
and the mean-field version \eq{eq13} of Ohm's law takes the form
\EQ
\bmJ = \sigma_{\rm m} (\bmE + \alpha \bmB)
\label{eq53}
\EN
with the mean-field conductivity $\sigma_{\rm m}$ given by
\EQ
\sigma_{\rm m} = \sigma / (1 + \mu \sigma \beta) \, .
\label{eq55}
\EN
While $\alpha$ is a pseudoscalar, $\beta$ is a scalar.

Homogeneity and isotropy of turbulence do not exclude reflexional symmetry.
We define it by the invariance of all averaged quantities depending on $\bu$ under reflexion  of the $\bu$ field at a point.
In the case of homogeneity and isotropy this is equivalent to reflexions at any plane.
A reflexion turns a right-handed structure in the flow field in a left-handed one and vice versa.
Reflexional symmetry in the above sense implies therefore an equipartition of right-handed
and left-handed structures in a fluid flow, that is, the absence of any preferred handedness.
A simple consequence is, e.g., that the mean kinetic helicity $\ol{\bu \cdot (\nab \x \bu)}$ vanishes.
In the case of homogeneous isotropic ref
lexionally symmetric turbulence we may easily show
that the pseudoscalar $\alpha$ in \eq{eq51} and \eq{eq53} has to be zero.
There is, however, no restriction to the scalar $\beta$.

As long as there are no deviations of the underlying homogeneous isotropic turbulence from reflexional symmetry
the mean-field version of Ohm's law reads simply $\bmJ = \sigma_{\rm m} \bmE$ with $\sigma_{\rm m}$ as given by \eq{eq55}.
The insight, that for mean fields a conductivity different from that relevant for the original fields applies,
can already be found in papers by Sweet (1950) and Csada (1951).
As we will see later (section \ref{mfcoeff}) the ratio $\sigma_{\rm m} / \sigma$ can be much bigger than unity.
In the solar convection zone, e.g., it may take values of the order of $10^4$,
what explains in particular the observed life times of sunspots.

Turbulence in rotating bodies, on which we want to focus our attention later, is subject to the Coriolis force.
It deviates therefore not only from isotropy but also from reflexional symmetry.
(This corresponds to the fact that the Coriolis force is determined, e.g., by a right-hand rule.)
Preparing investigations of this complex situation, we consider here first the more or less academic case
of homogeneous isotropic but not reflexionally symmetric turbulence, in which \eq{eq51} and \eq{eq53}
with $\alpha \not= 0$ apply.
The occurrence of the electromotive force $\alpha \bmB$ is called ``$\alpha$ effect".
It has been first considered by Steenbeck, Krause and R\"adler (1966) in a slightly different context (see section \ref{realistic}).

The $\alpha$ effect allows growing mean magnetic fields, that is, dynamo action.
In order to show this we consider the mean-field induction equation \eq{eq15} with $\bmU = \bzo$
and $\bscE$ specified according to \eq{eq51}, that is,
\EQ
\eta_{\rm m} \nab^2 \bmB + \alpha \nab \x \bmB - \p_t \bmB = \bzo \, , \quad  \nab \cdot \bmB = 0 \, ,
\label{eq57}
\EN
where $\eta_{\rm m}$ is the magnetic mean-field diffusivity,
\EQ
\eta_{\rm m} = \eta + \beta \, .
\label{eq59}
\EN
Simple particular solutions $\bmB$ of \eq{eq57} read
\EQ
\begin{aligned}
\!\!\!\!\!\!\!\bmB &= B_0 (\cos kz , \pm \sin kz , 0 ) \exp(\lambda t) \, , \\
&\qquad \qquad \qquad \quad \lambda = - (\eta + \beta) k^2 \pm \alpha k \, ,
\label{eq61}
\end{aligned}
\EN
with a wave number $k$, which we consider as positive, and a growth rate $\lambda$.
Growing solutions, i.e., such with $\lambda > 0$, are possible as soon as $|\alpha| > (\eta + \beta) \, k$.
We will see later (section \ref{mfcoeff}) that this condition can well be satisfied.

\subsection{A more realistic example}
\label{realistic}

In a next, somewhat more realistic case, we consider turbulence on a rotating body.
We assume that for a co-rotating observer no mean flow exists, $\bmU = \bzo$, but admit slight deviations
of the turbulence, $\bu$, from homogeneity and isotropy.
We further assume that the inhomogeneity can be described by a vector $\bg$, e.g., the  intensity gradient $\nab \ol{u^2}$ of the turbulence.
The anisotropy depends, of course, apart from $\bg$ also on the angular velocity $\bOmega$ which defines the Coriolis force.
Again, we restrict ourselves to sufficiently small variations of the mean magnetic field $\bmB$ in space and time
so that \eq{eq35} applies.
Considering the deviations of the turbulence from homogeneity and isotropy as sufficiently small,
we assume that $a_{ij}$ and $b_{ijk}$ are linear in both $\bg$ and $\bOmega$.
Studying then the possible tensorial structures of $a_{ij}$ and $b_{ijk}$ we find
\EQA
&& \!\!\!\!\!\!\!\!\!\!\!\!\!\!\! \bscE =
\alpha_1 (\bg \cdot \bOmega) \bmB + \alpha_2 \bg (\bOmega \cdot \bmB) + \alpha_3 \bOmega (\bg \cdot \bmB)
    + \gamma \bg \x \bmB
\nonumber\\
&& \!\! - \beta \nab \x \bmB - \delta \bOmega \x (\nab \x \bmB) - \delta_* \nab (\bOmega \cdot \bmB)
\label{eq65}
\ENA
with scalar coefficients $\alpha_1$, $\alpha_2$, ... $\delta_*$ independent of $\bg$ and $\bOmega$.

The first line of \eq{eq65} reproduces the momentous result by Steenbeck, Krause and R\"adler (1966).
Due to the Coriolis force there is now (at least locally) a preference of either right-handed or left-handed helical patterns
in the turbulent flow.
As a consequence the three contributions to $\bscE$ with the coefficients $\alpha_1$, $\alpha_2$ and $\alpha_3$ occur in \eq{eq65}.
The term $\alpha_1 (\bg \cdot \bOmega) \bmB$ corresponds to $\alpha \bmB$ in relation \eq{eq51}, that is, in the case
of homogeneous isotropic turbulence lacking reflexional symmetry.
However, the $\alpha_1$ term is now accompanied by two others, the $\alpha_2$ and $\alpha_3$ terms.
We speak here again of an $\alpha$ effect, more precisely, if all three terms are considered, of an anisotropic $\alpha$ effect.
On a spherical body, on which $\bg$ points in radial direction, $\alpha_1 (\bg \cdot \bOmega)$ changes its sign
when moving from the northern hemisphere to the southern one.
The $\alpha$ effect as considered here is crucial for special types of mean-field dynamo mechanisms (see section \ref{dynmech}).

The $\gamma$ term in \eq{eq65} describes the transport of mean magnetic flux by inhomogeneous turbulence.
This effect has been first, for a two-dimensional turbulence, considered by Zeldovich (1956),
later in a more general context by R\"adler (1968a,b).
It has been discussed as ``pumping of mean magnetic flux" or (since the mean magnetic flux is expelled
from regions of high turbulence intensity) as ``turbulent diamagnetism".

The $\beta$ term corresponds to that which occurred already in the case of homogeneous isotropic turbulence, that is in \eq{eq51},
and gives rise to introduce the mean-field conductivity in the mean-field version of Ohm's law
or the mean-field diffusivity in the mean-field induction equation.

The effect described by the $\delta$ term in \eq{eq65} has been first considered by R\"adler (1969a,b).
It is often called ``$\bOmega \x \bJ$ effect", in what follows also simply ``$\delta$ effect".
Combined with mean shear it is capable of dynamo action (see section \ref{dynmech}).
Other than the $\alpha$ effect, the $\delta$ effect requires spatial variations of the mean magnetic field;
it does not occur with a homogeneous field.
Apart from this it is in a sense simpler than the $\alpha$ effect.
It needs no gradient of the turbulence intensity but occurs already with homogeneous turbulence.
The sum of the $\beta$ and $\delta$ terms can also be described by an anisotropic mean-field conductivity
with a non-symmetric conductivity tensor.
The $\delta_*$ term is of minor importance.
It does not influence the mean-field induction equation as long as $\delta_*$ is spatially constant.

\subsection{Mean-field dynamo mechanisms}
\label{dynmech}

When discussing dynamo mechanisms due to turbulent motions we focus attention on mean-field dynamos.
They are characterized by the ability to generate magnetic fields
with length scales much larger than the typical length scales of the turbulent motions.
Therefore we call them also ``large-scale dynamos".
In this context we should have in mind the finding by Kazantsev (1968) according to which
a homogenous isotropic turbulence, which needs not to deviate from reflexional symmetry, may under certain conditions
maintain irregular magnetic fields with length scales smaller than or equal to those of the turbulent motion,
which therefore do not contribute to a mean magnetic field.
We speak then of a ``small-scale dynamo".
The influence of large-scale on small-scale dynamos and vice versa is an interesting subject (see, e.g., Brandenburg et al. 2012),
which we however do not want to discuss here.

Let us consider dynamos due to turbulence on an axisymmetric rotating fluid body.
We restrict attention first to mean magnetic fields that are symmetric about the axis of rotation.
Each such field can be split into a poloidal part that lies completely in meridional planes,
and a toroidal part perpendicular to them.
Within this frame, dynamos can always be understood as an interplay between the poloidal and the toroidal part
of the mean magnetic field.
We admit here a mean motion in the form of differential rotation, that is, a dependence
of the corresponding angular velocity $\Omega$ on radius or latitude.
In addition to induction effects of turbulent motions described, e.g., by $\alpha$ or $\delta$ effects,
we have then also the effect of rotational shear, which we call ``$\Omega$ effect".
While $\alpha$ and $\delta$ effect generate poloidal magnetic fields from toroidal ones and vice versa,
the $\Omega$ effect generates only a toroidal field from a poloidal one.

The simplest mean-field dynamo mechanism is that of $\alpha^2$ type, in which both the generation
of the poloidal field from the toroidal one as well as that of the toroidal field from the poloidal one
is due the $\alpha$ effect.
The first spherical models of this type were proposed by Steenbeck and Krause (1969b), and were on the basis
of numerical simulations discussed in view of the Earth and the planets.
If we admit in addition the $\Omega$ effect and assume that it dominates the generation of the toroidal field,
we arrive at dynamos of the $\alpha \Omega$ type.
Models of this type have been investigated also by Steenbeck and Krause (1969a) and applied to explain essential features
of the magnetic solar cycle.
The case in which both the $\alpha$ and the $\Omega$ effect contribute markedly to the generation of the toroidal field
is often labeled as mechanism of $\alpha^2 \Omega$ type.
In the course of time, a large number of dynamo models of the mentioned types have been investigated (see, e.g., R\"adler 1986).

Mean-field dynamos require not necessarily the $\alpha$ effect.
It is easy to see that there is no dynamo due to the $\delta$ effect alone.
However, the combination of the $\delta$ and $\Omega$ effects is, as demonstrated by R\"adler (1969b, 1976), capable of dynamo action.
Meanwhile there are several results for dynamo models of that type (see, e.g, R\"adler 1974, 1986).
Also the combination of another effect which can be described as an anisotropy of the mean-field conductivity
with the $\Omega$ effect can lead to a dynamo (R\"adler 1986).

So far we focussed attention on dynamos with axisymmetric mean fields, which allow a simple description.
The mechanisms mentioned here work, however, also with non-axisymmetric mean fields, and there are quite a few cases
in which such fields are easier to excite than axisymmetric ones (see, e.g., R\"adler 1986).

Rogachevskii and Kleorin (2004) claimed that the induction effects that occur in a turbulent fluid
under the influence of a global shear, which are similar to those in the $\delta \Omega$ mechanism,
are also capable of dynamo action, and they spoke of a ``shear-current dynamo".
They calculated the relevant mean-field coefficients however in a defeasible approximation.
Several investigations on this "shear-current dynamo" have been carried out, but no reliable proof for its existence
has been given so far.

It has been shown for a wide range of assumptions that the magnetic mean-field diffusivity $\eta_{\rm m} = \eta + \beta$
is positive so that a growth of a mean magnetic field due to negative mean diffusivity can be excluded.
There is however a very recent result which might provoke scruples in this respect.
Some properties of mean-field dynamos are reflected by mean-field models derived from the dynamos investigated by Roberts (1972),
working not with turbulence but with regular three-dimensional flows periodic in, say, $x$ and $y$ and independent of $z$.
It has been shown very recently by Devlen et al. (2013) that in one of these mean-field models growing mean magnetic fields
are generated with no other mean-field induction effect than a negative mean-field diffusivity.
It remained open whether this result can be extended to mean-field dynamos working with real turbulence.

\subsection{Calculation of mean-field coefficients}
\label{mfcoeff}

\subsubsection{Second-order correlation approximation}
\label{soca}

So far we have have considered connections of the mean electromotive force $\bscE$ with the mean magnetic field $\bmB$
and its spatial derivatives that are defined by coefficients like $\alpha$, $\beta$, $\alpha_1$ or $\delta$,
but nothing has been said about their actual values or their dependence on the magnitude or other properties
of the fluctuating motions.
In the early days of mean-field electrodynamics many calculations of $\bscE$ were carried out on the basis
of equation \eq{eq19} for the magnetic fluctuations $\bb$ but with the $\bepsilon$ term canceled.
This is an approximation that can be justified for sufficiently small velocity fluctuations $\bu$ only,
often called ``second-order correlation approximation" (SOCA) or ``first-order smoothing approximation" (FOSA)
or ``quasilinear approximation".

Consider as an example again the case in which the mean velocity $\bmU$ vanishes and the fluctuating velocity $\bu$
corresponds to a homogeneous isotropic turbulence.
Restrict attention further to small variations of the mean magnetic field $\bmB$ in space and time,
that is, ideal scale separation, and to the in applications most important high-conductivity limit,
defined by $\eta \tau_{\rm c} / \lambda_{\rm c}^2 \ll 1$,
where $\tau_{\rm c}$ and $\lambda_{\rm c}$ are correlation time and correlation length of the turbulent motions.
Within the frame of SOCA we find then
\EQ
\begin{aligned}
\alpha &= - {\textstyle \frac{1}{3}} \int_0^\infty \ol{\bu (t) \cdot (\nab \x \bu (t - \tau))} \, \dd \tau \\
\beta &= {\textstyle \frac{1}{3}} \int_0^\infty \ol{\bu (t) \cdot \bu (t - \tau)} \, \dd \tau \, .
\end{aligned}
\label{eq101}
\EN
We may write this also in the form
\EQ
\alpha = - {\textstyle{1 \over 3}} \ol{\bu \cdot (\nab \x \bu)} \, \tau^{(\alpha)} \, , \quad \beta = {\textstyle{1 \over 3}} \ol{u^2} \, \tau^{(\beta)}
\label{eq103}
\EN
with $\tau^{(\alpha)}$ and $\tau^{(\beta)}$ defined by equating the corresponding expressions in \eq{eq101} and \eq{eq103}.
Under reasonable assumptions both $\tau^{(\alpha)}$ and $\tau^{(\beta)}$ are approximately equal to $\tau_{\rm c}$.
These results are in many respects instructive.
In the high-conductivity limit considered here, however, the application of SOCA can only readily be justified
if the Strouhal number $St = u_{\rm c} \tau_{\rm c} / \lambda_{\rm c}$,
with $u_{\rm c}$ being a characteristic value of $\bu$, is small compared with unity.
In realistic cases of turbulence it is close to unity.

It is well possible to proceed from the second-order approximation to a third-order one with $\bepsilon$
expressed by second-order results, then to a fourth-order one with $\bepsilon$ expressed by third-order results etc.,
and it has been proven that this procedure converges (Krause 1968).
Analytic calculations of that kind are however very tedious and, apart from a few fourth-order results,
no results of practical interest have been gained in this way.

\subsubsection{Test-field method}
\label{testfield}

Several other techniques for obtaining results for mean-field coefficients have been proposed, using assumptions which look
to a certain extent plausible but cannot be justified in a clean way
(for a critical review see, e.g., R\"adler and Rheinhardt 2007).
In recent years, with growing possibilities of numerical calculations, the ``test-field method", established
immediately on the basic equations, brought much progress in the reliable determination of mean-field coefficients.

The method was developed by Schrinner et al. (2005, 2007) in the context of this task:
Consider a simple geodynamo model, with the magnetic field maintained by convection.
Define mean fields by averaging over the azimuthal coordinate;
they are then axisymmetric.
Extract the mean-field coefficients from the numerical results for this model.
Construct a mean-field model with these coefficients.
Compare then the mean fields obtained from the original model by direct numerical simulations
with those obtained from the mean-field model.
In the ideal case they should agree with each other.

Let us sketch the idea of the test-field method for the case of the simple connection of the mean electromotive force $\bscE$
with $\bmB$ and its first spatial derivatives as given by \eq{eq35}.
We choose a set of test fields $\bmB^{\rm T}$
and replace $\bmB$ in \eq{eq19} consecutively by each of its elements, calculate the corresponding $\bb^{\rm T}$
and finally $\bscE^{\rm T} = \ol{\bu \x \bb^{\rm T}}$.
These $\bscE^{\rm T}$ have to obey
\EQ
a_{ij} \mB_j^{\rm T} + b_{ijk} \p \mB_j^{\rm T} / \p x_k = {\cE}_i^{\rm T} \, .
\label{eq105}
\EN
With a sufficient number of independent $\bmB^{\rm T}$ we obtain a system of equations which allows us the determination
of the $a_{ij}$ and $b_{ijk}$ from the ${\cE}_i^{\rm T}$ calculated for the chosen set of $\bmB^{\rm T}$.
It turned out that there is a high degree of freedom in the choice of the test fields.
They need not to be solenoidal and have not to satisfy specific boundary conditions.

\begin{figure}\begin{center}
\includegraphics[width=0.9\columnwidth]{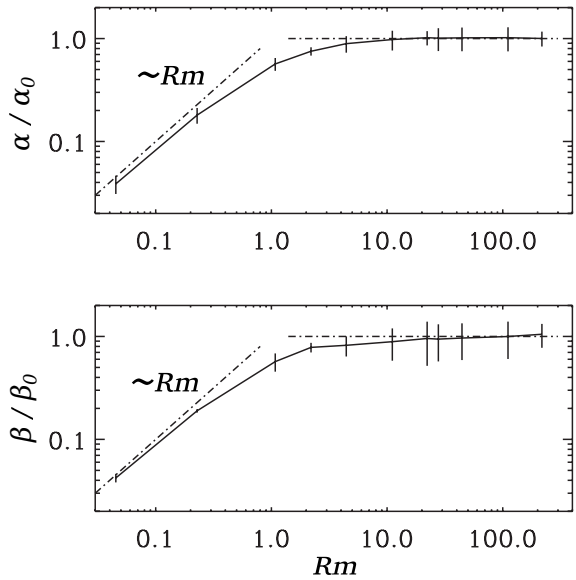}
\end{center}\caption[]{
Normalized mean-field coefficients $\alpha/\alpha_0$ and $\beta/\beta_0$ as functions of $Rm$,
obtained in test-field calculations by Sur et al. (2009) based on turbulence simulations with $Re = 2.2$
}\label{fig1}\end{figure}

\begin{figure}\begin{center}
\includegraphics[width=0.95\columnwidth]{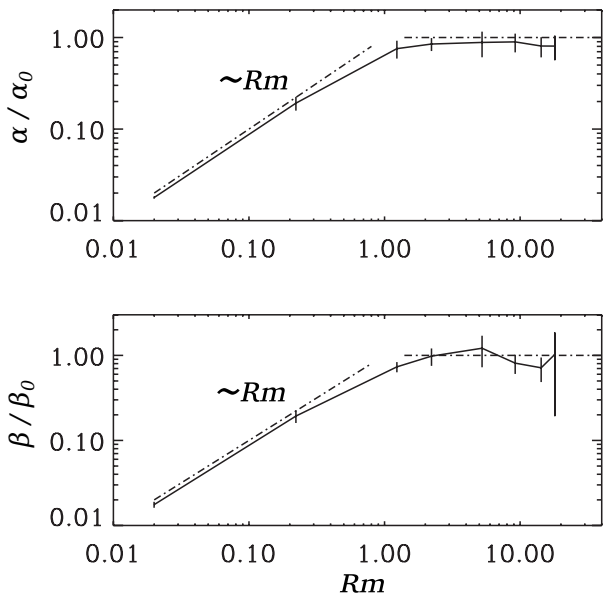}
\end{center}\caption[]{
Same as Fig.1 but simulations with $Re = 10 \, Rm$
}\label{fig2}\end{figure}

Let us return once more to the coefficients $\alpha$ and $\beta$ for homogeneous isotropic turbulence.
Referring to numerical simulations of hydrodynamic turbulence in a weakly compressible fluid,
Sur et al. (2008) used the test-field method for the determination of these coefficients.
The turbulence was specified to have an energy input at a wavenumber $k_{\rm f}$, and to be maximally helical,
that is, $\ol{(\nab \x \bu)^2} / \ol{\bu^2} = k_{\rm f}^2$.
Calculations with different values of the hydrodynamic Reynolds number $Re = u_{\rm rms} / \nu k_{\rm f}$,
where $\nu$ means the kinematic viscosity, were carried out.
In Figs.\ref{fig1} and \ref{fig2} some results for $\alpha / \alpha_0$ and $\beta / \beta_0$ with
$\alpha_0 = - {\textstyle {1 \over 3}} u_{\rm rms}$ and $\beta_0 = {\textstyle {1 \over 3}} u_{\rm rms} / k_{\rm f}$
are shown in dependence on the magnetic Reynolds number $Rm = u_{\rm rms} / \eta k_{\rm f}$.
In the turbulence considered here the Strouhal number $St$ turned out to be of the order of unity.
So the reported results confirm that \eq{eq101} and \eq{eq103}, which were derived for $St \ll 1$ only, apply also
with realistic values of $St$.

The test-field method for the determination of the mean-field coefficients brought much progress
in mean-field electrodynamics and beyond.
It has been extended to a very broad range of assumptions, is in particular not limited to cases with scale separation
(see, e.g., Brandenburg et al. 2008, Rheinhardt and Brandenburg 2010, 2012).

\subsection{Imperfect scale separation}
\label{imperfect}

\subsubsection{Apparent discrepancies}
\label{nonlocal}

In the examples considered so far we have reduced the general representations \eq{eq25} or \eq{eq31}
of the mean electromotive force $\bscE$ as a convolution depending on the mean magnetic field $\bmB$
in all space and at the current time and all past times,
to the simple local and instantaneous connection \eq{eq35} of $\bscE$ with $\bmB$ and its first spatial derivatives.
On this level the theory may deliver incomplete or even wrong statements.

One example for that is the aforementioned incomplete agreement of the mean-field geodynamo models
derived immediately from the basic equations and those constructed with mean-field coefficients determined by the simple version
of the test-field method, which considers only local and instantaneous connections of $\bscE$ with $\bmB$ and its first derivatives
(section \ref{testfield}).

\subsubsection{Memory effect}
\label{memory}

Another interesting example concerns the growth of a mean magnetic field in a turbulence showing $\alpha$ effect.
As Hubbard and Brandenburg (2009) pointed out, the growth rates obtained in direct numerical simulations clearly differ
from those derived from a dispersion relation with mean-field coefficients gained in a static approximation,
that is, assuming an instantaneous connection of $\bscE$ and $\bmB$ as in \eq{eq35}.
The difference disappears if a proper connection of $\bscE$ at a given time with $\bmB$ at former times, that is,
some memory of the turbulent system, is taken into account.
We know meanwhile many examples in which such memory effects play an important role and can even be crucial
for the existence of dynamos (Rheinhardt et al. 2014).

\begin{figure}\begin{center}
\includegraphics[width=0.89\columnwidth]{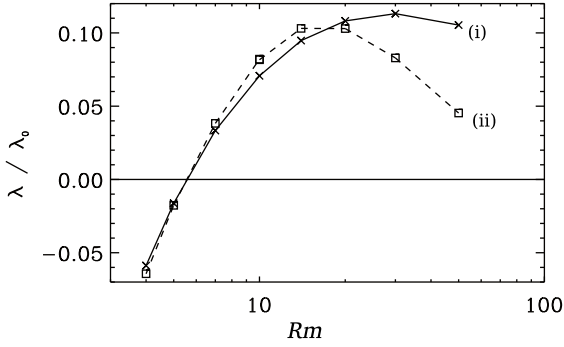}
\end{center}\caption[]{
Normalized growth rates $\lambda / \lambda_0$ of a mean magnetic field in a Roberts flow as functions of $Rm$,
(i) obtained in direct numerical simulations and (ii) calculated from a dispersion relation with mean-field coefficients
determined in a static approximation, according to Hubbard and Brandenburg (2009)
}\label{fig3}\end{figure}

For an illustration of the memory effect, Hubbard and Brandenburg (2009) considered a Roberts flow instead of a real turbulence.
They assumed $\bu \!=\!- \be \x \nab \psi + k_{\rm f} \psi \,\be$ and $\psi = (u_0 / k_0) \cos k_0x \cos k_0y$,
where $\be$ means the unit vector in $z$ direction and $u_0$, $k_{\rm f}$ and $k_0$ are constants,
and they restricted attention on the case of a maximal modulus of the relative helicity
$\ol{\bu \cdot (\nab \x \bu)} / \ol{\bu^2} k_{\rm f}$, which occurs with $k_{\rm f} = \sqrt{2} k_0$.
They further defined mean fields by averaging over all $x$ and $y$.
Fig~\ref{fig3} shows the normalized growth rates $\lambda / \lambda_0$, with $\lambda_0 = u_{\rm rms} k_{\rm f}$,
as functions of $Rm$, obtained (i) in direct numerical simulations and (ii) from the dispersion relation
with mean-field coefficients determined in a static approximation.
Note the substantial deviations of the two results for large $Rm$.

\section{Mean-field magnetohydrodynamics}

\subsection{Momentum balance and consequences}

So far the fluid velocity has been considered as prescribed.
We now relax this assumption and use in addition to the electromagnetic equations \eq{eq01} and \eq{eq03},
or the induction equation \eq{eq05}, also the momentum balance.
For the sake of simplicity we restrict ourselves to an incompressible fluid.
Admitting a rotating frame of reference we have then
\EQA
&&\!\!\!\!\!\!\!\!\!\!\!\! \varrho (\p_t \bU + (\bU \cdot \nab) \bU) = - \nab P + \varrho \nu \nab^2 \bU - 2 \varrho \bOmega \x \bU
\nonumber\\
&& \quad + (1/\mu) (\nab \x \bB) \x \bB + \bF \, , \quad \nab \cdot \bU = 0 \, .
\label{eq111}
\ENA
Here $\varrho$ means the mass density, $\nu$ again the kinematic viscosity of the fluid, and $P$ the hydrodynamic pressure.
The angular velocity $\bOmega$ defines the rotation of the frame and so the Coriolis force,
and $\bF$ stands for any external force.
The inertial term in \eq{eq111} is balanced by the pressure gradient, the viscous force, the Coriolis force, the Lorentz force
and possibly some external force.

Let us focus attention again on turbulent situations.
Taking then the average not only of the induction equation \eq{eq05} but also of the momentum balance \eq{eq111},
we find in addition to the mean-field induction equation \eq{eq15} with the mean electromotive force $\bscE$ given by \eq{eq17}
the mean-field version of the momentum balance,
\EQA
&&\!\!\!\!\!\!\!\!\!\!\!\! \varrho (\p_t \bmU + (\bmU \cdot \nab) \bmU) = - \nab \ol{P} + \varrho \nu \nab^2 \bmU
    - 2 \varrho \bOmega \x \bmU
\nonumber\\
&&\!\!\!\!\!\!\! + (1/\mu) (\nab \x \bmB) \x \bmB + \bmF + \bscF  \, , \quad \nab \cdot \bmU = 0 \, ,
\label{eq113}
\ENA
with a mean ponderomotive force $\bscF$,
\EQ
\bscF = - \varrho \ol{(\bu \cdot \nab) \bu} + (1/\mu) \ol{(\nab \x \bb) \x \bb} \, .
\label{eq115}
\EN

If we ignore the magnetic field we return to pure hydrodynamics.
The mean ponderomotive force $\bscF$ covers then, for example, the contribution of the turbulence
to the mean-field viscosity, often discussed as eddy viscosity, further the $\Lambda$ effect, which,
on a rotating body, may drive differential rotation (R\"udiger 1989), or the anisotropic kinetic $\alpha$ effect
(AKA effect, Frisch et al. 1987).
We do not want to discuss these subjects here but focus attention on cases with magnetic field.

Generalizing the considerations on the mean electromotive force $\bscE$ explained above,
we see that both the electromotive force $\bscE$ and the ponderomotive force $\bscF$ may be considered as functionals
of fluctuations $\bu^{(0)}$ and $\bb^{(0)}$, for example relating to the case of vanishing mean motion and mean magnetic field,
of the mean velocity $\bmU$, the mean magnetic field $\bmB$ and their first spatial derivatives
and also of the angular velocity $\bOmega$ that determines the Coriolis force.
These functionals are not necessarily linear in $\bmU$, $\bmB$ or $\bOmega$.

Focussing attention on the mean electromotive force $\bscE$ we write once again $\bscE = \bscE^{(0)} + \bscE^{(B)}$,
with a part $\bscE^{(0)}$ independent of the mean magnetic field $\bmB$.
In our short presentation of mean-field electrodynamics we argued, considering purely hydrodynamic background turbulence,
that the part $\bscE^{(0)}$ will always decay to zero.
Now we may no longer exclude magnetohydrodynamic turbulence, and then this is no longer necessarily the case.
A non-zero part $\bscE^{(0)}$ of $\bscE$ corresponds to a battery.
If such a part exists, the mean-field induction equation is no longer homogeneous in the mean magnetic field $\bmB$
and has always non-decaying solutions.
In the absence of conditions that allow a mean-field dynamo the magnitude of the corresponding mean magnetic fields
is determined by $\bscE^{(0)}$.
If a dynamo is possible, they may act as seed fields.

\subsection{Yoshizawa effect}

An interesting example of a contribution to the mean electromotive force independent of the mean magnetic field, $\bscE^{(0)}$,
has been given by Yoshizawa in 1990.
Let us consider magnetohydrodynamic turbulence in the presence of a mean flow or on a rotating body,
that is, under the influence of the Coriolis force.
Assuming originally homogeneous isotropic turbulence and ideal scale separation in space and time,
we may expect that
\EQ
\bscE^{(0)} = c_U \bmU + c_W \nab \x \bmU + c_\Omega \bOmega
\label{eq121}
\EN
with three coefficients $c_U$, $c_W$ and $c_\Omega$.
If the turbulence shows Galilean invariance, $c_U$ has to be zero.
The two coefficients $c_W$ and $c_\Omega$ must be pseudo scalars, and it turns out that they are closely connected
with the cross-helicity $\ol{\bu \cdot \bb}$.
The Yoshizawa effect, that is, an electromotive force like \eq{eq121} with nonzero $c_W$ or $c_\Omega$,
is capable of building up and maintaining a mean magnetic field.
Its strength depends on $c_W$ and $c_\Omega$.
As explained above, it may act as a seed field if there are conditions which allow a further growth of mean magnetic fields.
Of course, the production of cross-helicity in general depends also on the mean magnetic field,
and in that sense the mean electromotive force under consideration, too, may depend on this mean magnetic field.

\subsection{$\alpha$ effect and $\alpha$ quenching}

Let us return to the situation as considered in section \ref{simple}, that is, no mean motion, no Coriolis force
and only small variations of $\bmB$ in space and time.
Instead of purely hydrodynamic turbulence we assume however now homogeneous isotropic magnetohydrodynamic turbulence,
for which $\bb$, like $\bu$, remains non-zero if $\bmB \to \bzo$.
Then $\bscE$ has to satisfy again \eq{eq51}, that is $\bscE = \alpha \, \bmB - \beta \, (\nab \x \bmB)$.
Solving the equation governing $\bb$ under SOCA and that for $\bu$ under an analogous approximation,
further restricting ourselves to the high-conductivity limit, $\eta \tau_{\rm c} / \lambda_{\rm c}^2 \ll 1$,
and the analogous low-viscosity limit, $\nu \tau_{\rm c} / \lambda_{\rm c}^2 \ll 1$, we find
\EQA
\alpha &=& \alpha_{\rm K} + \alpha_{\rm M}
\nonumber\\
\alpha_{\rm K} &=& - {\textstyle{1 \over 3}} \ol{\bu^{(0)} \cdot (\nab \x \bu^{(0)})} \, \tau^{(\alpha {\rm K})}
\label{eq125}\\
\alpha_{\rm M} &=&  {\textstyle{1 \over 3 \mu \varrho}} \ol{\bb^{(0)} \cdot (\nab \x \bb^{(0)})} \, \tau^{(\alpha {\rm M})}
\nonumber
\ENA
and
\EQ
\beta = \beta_{\rm K} = {\textstyle{1 \over 3}} \ol{{\bu^{(0)}}^2} \, \tau^{(\beta)} \, .
\label{eq127}
\EN
(see, e.g., R\"adler and Rheinhardt 2007).
Here $\bu^{(0)}$ and $\bb^{(0)}$ stand for $\bu$ and $\bb$ in the limit $\bmB \to \bzo$,
and $\tau^{(\alpha {\rm K})}$, $\tau^{(\alpha {\rm M})}$ and $\tau^{(\beta)}$ are quantities approximately equal
to the correlation time $\tau_{\rm c}$.
Within this framework the $\alpha$ effect has in addition to the kinetic part, which is connected
with the mean kinetic helicity $\ol{\bu \cdot (\nab \x \bu)}$, a magnetic part connected
with the mean current helicity $\ol{\bb \cdot (\nab \x \bb)} = \ol{\mu \bj \cdot \bb}$.
Such a magnetic part has been first considered by Pouquet et al. in 1976.
Remarkably the coefficient $\beta$, which determines the mean-field diffusivity, has no such magnetic contribution.
If we put $\bb^{(0)} = 0$ we return to our old result for purely hydrodynamic turbulence.

Let us now admit an arbitrarily strong mean magnetic field.
It causes an anisotropy of the turbulence such that the tensor $a_{ij}$ in \eq{eq35}
has the structure $\alpha_1 \delta_{ij} + \alpha_2 e_i e_j$, where $\alpha_1$ and $\alpha_2$ may depend on $|\bmB|$,
and $\be$ stands for the unit vector in the direction of $\bmB$.
Considering then \eq{eq35} but ignoring, for simplicity, the terms with derivatives of $\bmB$, we find again $\bscE = \alpha \, \bmB $
with $\alpha = \alpha_1 + \alpha_2$ being a function of $|\bmB|$.
In general we expect a reduction of the modulus of $\alpha$ with growing $|\bmB|$.
In this case we speak of ``$\alpha$ quenching".
It limits the growth of the mean magnetic field and defines a saturation field strength.

The determination of the dependence of $\alpha$ on $|\bmB|$ is a complex task.
A simple ansatz that has been frequently discussed in the past reads
\EQ
\alpha = \frac{\alpha_0}{1 + c \, \mB^2 / B^2_{eq}} \, ,
\label{eq131}
\EN
where $\alpha_0$ is the value of $\alpha$ in the limit $\bmB \to \bzo$, further $c$ a dimensionless positive constant
and $B_{eq}$ the equipartition field strength defined such that the energies stored in the fluctuating velocity field
and in the mean magnetic field are equal to each other, that is, $B^2_{eq} = \mu \varrho \, \ol{u^2}$.

In 1992 Vainshtein and Cattaneo suggested on the basis of analytical considerations and numerical calculations
with an imposed magnetic field a relation like \eq{eq131} with $c \approx Rm$,
where $Rm$ means again the magnetic Reynolds number.
In the solar convection zone, for example, $Rm$ takes values of $10^6$ or even $10^9$,
and $\mB / B_{eq}$ values of the order of unity.
Then $\alpha$ would be very close to zero and we could not expect any dynamo.
Therefore this kind of quenching has been called ``catastrophic quenching".

This finding has initiated many discussions and investigations.
Considerable progress has been made by investigating the simplest possible dynamo systems with $\alpha$ effect
in the nonlinear regime.
A fully satisfactory theory of this subject is, however, still missing.

One important issue in the recent investigations on $\alpha$ quenching are the hypotheses that $\alpha$ is always a sum
of a kinetic part $\alpha_{\rm K}$ and a magnetic part $\alpha_{\rm M}$
and that the latter is determined by the part of the mean current helicity due to the electric current and magnetic field fluctuations,
$\ol{\bj \cdot \bb}$.
The other important issue is the role of the magnetic helicity in a dynamo.
We recall here that the magnetic helicity, say $H$, is defined as a volume integral over the magnetic helicity density
$h = \bA \cdot \bB$, where $\bA$ is a vector potential of the magnetic field $\bB$, that is $\nab \x \bA = \bB$.
If the electromagnetic fields satisfy specific conditions at the surface of this volume,
in particular the magnetic field does not intersect this surface, $H$ is independent
of the special choice of the vector potential $\bA$, that is, under gauge transformations of $\bA$.
Then the basic equations imply further that, in the limit of infinite conductivity, $H$ is a conserved quantity,
that is, does not change in time.
Within the mean-field concept the magnetic helicity density $h$ is the sum of two parts,
one originating from the mean magnetic field $\bmB$
and the other from the fluctuating part of the magnetic field, $\bb$.
The mean part of the latter, $\ol{\ba \cdot \bb}$ with $\nab \x \ba = \bb$, is closely related
to the magnetic contribution $\alpha_{\rm M}$ to $\alpha$, which is, as explained above, determined by the part
$\ol{\bj \cdot \bb} = (1/\mu) \ol{\bb \cdot (\nab \x \bb)}$ of the mean current helicity.
If, for example, in the limit of infinite conductivity $H$ is initially equal to zero and the mean magnetic field grows,
the modulus of $\alpha_{\rm M}$ must grow, too.

With the hypothesis $\alpha =  \alpha_{\rm K} + \alpha_{\rm M}$, further the evolution equation
of the mean magnetic helicity density due to fluctuations, and a few plausible assumptions an evolution equation for $\alpha_{\rm M}$,
\EQ
\p_t \alpha_{\rm M} = - 2 \eta_{\rm t} k_{\rm f}^2 \Big( \frac{\bscE \cdot \bmB}{B_{eq}^2} + \frac{\alpha_{\rm M}}{Rm} \Big)
   - \nab \cdot \bF \, ,
\label{eq135}
\EN
has been derived (see, e.g., Hubbard and Brandenburg 2011).
As usual in this context, we write here $\eta_{\rm t}$ instead of $\beta$, and $k_{\rm f}$ denotes again the wavenumber
of the energy-carrying scale in the turbulence.
$\bscE$ should be specified to be equal to $(\alpha_{\rm K} + \alpha_{\rm M}) \bmB - \eta_t \nab \x \bmB$.
As above, $B_{eq}$ is the equipartition field strength, and $\bF$ means a mean magnetic helicity flux.
In simple models with periodic boundary conditions the term $\nab \cdot \bF$ does not change
the total mean magnetic helicity inside a dynamo volume.
In general, however, the mean magnetic helicity flux plays a crucial role,
and expressions for $\bF$ have been elaborated which depend, for example, on differential rotation.
Models incorporating such results reflect indeed many properties of dynamos in the non-linear regime
including saturation field strengths (see, e.g., Brandenburg and Subramanian 2005, Hubbard and Brandenburg 2011,2012,
Del Sordo et al. 2013).

\section{Laboratory experiments}

The development of dynamo theory was accompanied and supported by several laboratory experiments.
As early as in 1967, one year after the first paper about this subject, the $\alpha$ effect has been demonstrated
in a liquid sodium flow in the Institute of Physics in Riga (Steenbeck et al. 1967).
The measurements were carried out at the so-called ``$\alpha$ box", in which a proper flow geometry has been
organized by baffles.

Already at this time there were many discussions on the realization of a dynamo in a conducting fluid.
It was clear from the very beginning that such an experiment requires a large fluid volume and high flow rates.
Only in the last days of the last century, in December 1999, after expensive preparations, two dynamos ran successfully
with liquid sodium flows, one in Riga (Gailitis et al. 2000) and one in Karlsruhe (M\"uller and Stieglitz 2000, 2002).
The first one (Riga) is clearly different from a mean-field dynamo, but the second one (Karlsruhe) can be well understood
as a mean-field dynamo of $\alpha^2$ type (see R\"adler et al. 2002).

I do not want to go into the details of these experiments but add a more general remark on the sometimes underestimated
practical value of basic research.
We have learned in geophysically or astrophysically motivated studies that the self-excitation of magnetic fields
in moving electrically conducting fluids is possible as soon as the magnetic Reynolds number $Rm = U L / \eta$,
with $U$ and $L$ being typical values of fluid velocity and linear dimensions of the considered device, exceeds a critical value,
which depends on the flow geometry and lies in all investigated cases above unity.
For a long time situations of that kind did not appear in laboratories or in industrial devices.
In the sixties and seventies of the last century, however, big fast breeder reactors were built with huge circuits of liquid sodium,
which transport the heat produced in the active zone to the places where it is transformed into electric power.
Such devices imply indeed the possibility of self-excitation of magnetic fields,
what constitutes a big danger.
These fields could quickly grow, heavily hamper the sodium flow and so seriously disturb the whole reactor regime
or even cause a catastrophy.
At first the reactor engineers were not aware of that.
It was the people thinking about cosmic dynamos who pointed out this danger.
Max Steenbeck, as a Foreign Member of the Soviet Academy, presented there in 1971 a corresponding memorandum
and initiated so investigations in this field and measures to avoid this danger.
Some years later, independent of that, two papers of British scientists about this topic have appeared (Bevir 1973, Pierson 1975).
This development fostered the willingness of authorities to support theoretical and experimental research on dynamos and related topics.

%%%%%%%%%%%%%%%%%%%%%%%%%%%%%%%%%%%%%%%%%%%%%%%%%%%%%%


\begin{thebibliography}{}

\bibitem[Babcock (1947)]{Bab47}
Babcock, H.W.: 1947, ApJ, 105, 105

\bibitem[Bevir (1973)]{Bev73}
Bevir, M.K.: 1973, Journ. British Nuclear Soc., 121, 455

\bibitem[Braginsky(1964)]{Brag54}
Braginsky, S.I.: 1964, Sov. Phys. JETP 20, 1462

\bibitem[Brandenburg(2005)]{Bra05}
Brandenburg, A., Subramanian, K.: 2005, Phys. Rep., 417, 1

\bibitem[Brandenburg(2008)]{Bra08}
Brandenburg, A., R\"adler, K.-H., Schrinner, M.: 2008, A\&A, 482, 739

\bibitem[Brandenburg(2012)]{Bra12}
Brandenburg, A., Sokoloff, D., Subramanian, K.: 2012, Space Sci. Rev., 169, 123

\bibitem[Bullard(1954)]{Bul54}
Bullard, E.C., Gellman, H.: 1954, Phil. Trans. R. Soc. A 247, 213

\bibitem[Cowling(1934)]{Cow34}
Cowling, T.G.: 1934, MNRAS 94, 39

\bibitem[Del Sordo (2013)]{DelSor13}
Del Sordo, F., Guerrero, G., Brandenburg, A.: 2013, MNRAS, 429, 1686

\bibitem[Devlen(2013)]{Dev13}
Devlen, E., Brandenburg, A., Mitra, D.: 2013, MNRAS 432, 1651

\bibitem[Elsassser(1946)]{Els58}
Elsasser, W.M.: 1946, Phys. Rev., 69, 106

\bibitem[Frisch(1987)]{Fri87}
Frisch, U., She, Z.S., Sulem, P.L.: 1987, Physica, 28, 382

\bibitem[Gailitis(2000)]{Gai00}
Gailitis, A., Lielausis, O., Dementev, S., Platacis, E., Cifersons, A., Gerbeth, G., Gundrum, T., Stefani, F.,
Christen, M., H\"anel, H., Will, G.: 2000, Phys. Rev. L., 84, 4365

\bibitem[Herzenberg (1958)]{Herz58}
Herzenberg, A.: 1958, Phil. Trans. R. Soc. London A, 250, 543

\bibitem[Hubbard(2009)]{Hub09}
Hubbard, A., Brandenburg, A.: 2009, ApJ, 706, 712

\bibitem[Hubbard(2011)]{Hub11}
Hubbard, A., Brandenburg, A.: 2011, ApJ, 727, 11

\bibitem[Hubbard(2012)]{Hub12}
Hubbard, A., Brandenburg, A.: 2012, ApJ 748, 51

\bibitem[Kazantsev(1968)]{Kaz68}
Kazantsev, A.P.: 1968, Sov. Phys. JETP, 16, 1031

\bibitem[Krause(1968)]{Kra68}
Krause, F.: 1968, ZAMM, 48, 333

\bibitem[Krause(1980)]{Kra80}
Krause, F., R\"adler, K.-H.: 1980, Mean-Field Magnetohydrodynamics and Dynamo Theory (Pergamon Press Oxford)

\bibitem[Larmor(1919)]{Larm1980}
Larmor, J.: 1919, Rep. Brit. Assoc. Adv. Sc. 1919, 159

\bibitem[Moffatt(1919)]{Moff78}
Moffat, H.K.: 1978, Magnetic Field Generation in Electrically Conducting Fluids (Cambridge University Press)

\bibitem[M\"uller (2000)]{Mue00}
M\"uller, U., Stieglitz, R.: 2000, Naturwissenschaften, 87, 381

\bibitem[M\"uller (2002)]{Mue02}
M\"uller, U., Stieglitz, R.: 2002, Nonlinear Processes in Geophysics, 9, 165

\bibitem[Parker (1955)]{Park55}
Parker, E.N.: 1955, Phil. Trans. R. Soc. London A, 250, 543

\bibitem[Parker (1957)]{Park57}
Parker, E.N.: 1957, Proc. N.A.S., 43, 8

\bibitem[Pierson (1975)]{Pier75}
Pierson, E.S.: 1975, Nuclear Science and Engineering, 57, 155

\bibitem[Pouquet (1976)]{Pou76}
Pouquet, A., Frisch, U., Leorat, J.: 1976, J. Fluid Mech., 77, 321

\bibitem[R\"adler (1968a)]{Rae68a}
R\"adler, K.-H.: 1968a, Z. Naturforsch. 23a, 1841

\bibitem[R\"adler (1968b)]{Rae68b}
R\"adler, K.-H.: 1968b, Z. Naturforsch. 23a, 1851

\bibitem[R\"adler (1969a)]{Rae69a}
R\"adler, K.-H.: 1969a, Monatsber. Dt. Akad. Wiss, 11, 194

\bibitem[R\"adler (1969b)]{Rae69b}
R\"adler, K.-H.: 1969b, Monatsber. Dt. Akad. Wiss, 11, 272

\bibitem[R\"adler(1976)]{Rae76}
R\"adler K.-H.: 1976, in: Bumba, V., Kleczek, J. (eds.),
Basic Mechanisms of Solar Activity, Proceedings IAU Symposium No. 71
(D. Reidel Publishing Company Dordrecht) p.323

\bibitem[R\"adler (1986)]{Rae86}
R\"adler, K.-H.: 1986, AN 307, 89

\bibitem[R\"adler et al. (2000)]{Rae00}
R\"adler, K.-H.: 2000, in: Page, D., Hirsch, J.G. (eds.), From the Sun to the Great Attractor (1999 Guanajuato Lecture in Astrophysics), p.101

\bibitem[R\"adler (2002)]{Rae02}
R\"adler, K.-H., Rheinhardt, R., Apstein, E., Fuchs, H.: 2002, Magnetohydrodynamics, 38, 39

\bibitem[R\"adler et al. (2007)]{Rae07}
R\"adler, K.-H., Rheinhardt, M.: 2007, Geophys. Astrophys. Fluid Dyn., 101, 117

\bibitem[Rheinhardt et al. (2010)]{Rhei10}
Rheinhardt, M., Brandenburg, A.: 2010, A\&A, 520, A28/1-16

\bibitem[Rheinhardt et al. (2012)]{Rhei12}
Rheinhardt, M., Brandenburg, A.: 2012, Astron.Nachr., 333, 71

\bibitem[Rheinhardt et al. (2014)]{Rhei14}
Rheinhardt, M., Devlen, E., R\"adler, K.-H., Brandenburg, A.: 2014, submitted to MNRAS, arXiv:1401.5026

\bibitem[R\"udiger (1989)]{Rue89}
R\"udiger, G.: 1989, Differential Rotation and Stellar Convection (Akademie-Verlag Berlin)

\bibitem[Schrinner et al. (2005)]{Schri05}
Schrinner, M., R\"adler, K.-H., Schmitt, D., Rheinhardt, M., Christensen, U.R.: 2005, AN 326, 245

\bibitem[Schrinner et al. (2007)]{Schri07}
Schrinner, M., R\"adler, K.-H., Schmitt, D., Rheinhardt, M., Christensen, U.R.: 2007, Geophys. Astrophys. Fluid Dyn., 101, 81

\bibitem[Steenbeck et al. (1966)]{Steen66}
Steenbeck, M., Krause, F., R\"adler, K.-H.: 1966, Z. Naturforsch. 21a, 369

\bibitem[Steenbeck et al. (1967)]{Steen67}
Steenbeck, M., Kirko, I.M., Gailitis, A., Klawina, A.P., Krause, F., Laumanis, I.J., Lielausis, O.A.: 1967, Mber. Dt. Akad. Wiss., 9, 714

\bibitem[Steenbeck et al. (1969a)]{Steen69a}
Steenbeck, M., Krause, F.: 1969a, AN 291, 271

\bibitem[Steenbeck et al. (1969b)]{Steen69b}
Steenbeck, M., Krause, F.: 1969b, AN, 291, 495

\bibitem[Sur et al. (2008)]{Sur08}
Sur, S., Brandenburg, A., Subramanian, K.: 2008, MNRAS, 385, l15

\bibitem[Sweet (1950)]{Sweet50}
Sweet, P.: 1950, MNRAS, 110, 69

\bibitem[Vainshtein (1992)]{Vain92}
Vainsthtein, S.I., Cattaneo, F.: 1992, ApJ, 393, 165

\bibitem[Yoshizawa (1990)]{Yosh90}
Yoshizawa, A.: 1990, PhFl B, 2, 1589

\bibitem[Zeldovich (1956)]{Zeld56}
Zeldovich, Y.B.: 1956, J. Exptl. Theoret. Phys., 31, 160


\end{thebibliography}
\end{document}